\documentclass[
reprint,
superscriptaddress,
showpacs,
 amsmath,amssymb,
 aps,
prl,
]{revtex4-1}

\usepackage{graphicx,epsf}
\usepackage{float}
\usepackage{xcolor}
\usepackage{psfrag}
\usepackage{ulem} 

\begin{document}

\title{Reversal of thermoelectric current in tubular nanowires}

\author{Sigurdur I.\ Erlingsson}
\affiliation{School of Science and Engineering, Reykjavik University, 
Menntavegur 1, IS-101 Reykjavik, Iceland}
\author{Andrei Manolescu}
\affiliation{School of Science and Engineering, Reykjavik University, 
Menntavegur 1, IS-101 Reykjavik, Iceland}
\author{George~Alexandru~Nemnes}
\affiliation{University of Bucharest, Faculty of Physics, MDEO 
Research Center, 077125 Magurele-Ilfov, Romania}
\affiliation{Horia Hulubei National Institute for Physics and 
Nuclear Engineering, 077126 Magurele-Ilfov, Romania}
\author{Jens H.\ Bardarson}
\affiliation{Max-Planck-Institut f\"ur Physik komplexer Systeme, 01187 Dresden, Germany}
\affiliation{Department of Physics, KTH Royal Institute of Technology, Stockholm, SE-106 91 Sweden}
\author{David Sanchez}
  \affiliation{Institute of Interdisciplinary Physics and Complex Systems IFISC (CSIC-UIB),
E-07122 Palma de Mallorca, Spain}


\begin{abstract} 
We calculate the charge current generated by a temperature bias between
the two ends of a tubular nanowire.  We show that in the presence of
a transversal magnetic field the current can change sign, i.e., electrons 
can either flow from the hot to the cold reservoir,  or in the opposite
direction, when the temperature bias increases. This behavior occurs when
the magnetic field is sufficiently strong, such that  Landau and snaking states are
created, and the energy dispersion is nonmonotonic with respect to the
longitudinal wave vector.  The sign reversal can survive in the presence
of impurities.  We predict this result for core/shell nanowires, for
uniform nanowires with surface states due to the Fermi level pinning,
and for topological insulator nanowires.
\end{abstract}


\maketitle

A temperature gradient across a conducing material induces an
energy gradient, which in turn results in particle transport.
In an open circuit, where no net current flows, a voltage is then
generated when two ends of a sample are maintained at different
temperatures --- this is the Seebeck effect and the linear voltage
response is known as thermopower.  The hotter particles have larger
average kinetic energy, and the net particle flow is therefore generally
from the hot to the cold side. The thermopower and
thermoelectric current can be positive or negative,
depending on the type of charge carriers, i.e., electrons or holes.

In comparison to this macroscopic case, the thermopower at the
nanoscale has special characteristics.  For example, if the energy
separation between the quantum states of the system is larger than
the thermal energy the thermopower may alternate between positive
and negative values, depending on the position of the Fermi level
relatively to a resonant energy, which can be controlled with a
gate voltage.  These oscillations were predicted a long time ago
\cite{Beenakker92}, and subsequently experimentally observed in
quantum dots \cite{Staring93,Dzurak93,Svensson12},  and in molecules
\cite{Reddy1568}.  A sign change in the thermopower can also be
obtained by increasing the temperature gradient and thus the population of
the resonant level \cite{Svensson13,Sierra14,Stanciu15,Zimbovskaya15}.  
In these examples the charge carriers are electrons and the sign change 
of the thermopower means that they travel from the cold side to the hot side,
which may appear counterintuitive.
Other nonlinear effects can occur if the characteristic
relaxation length of electrons and or phonons exceeds the sample size
\cite{Sanchez16}, because the energy of electrons and/or phonons is
no longer controlled by the temperature of the bath, but by the generated
electric bias, including Coulomb interactions \cite{Torfason13,Sierra16}.

Observing such negative thermopower at the nanoscale is difficult for at least two reasons: the currents tend to be small and it is hard to maintain a constant temperature difference across such short distances.
Here we argue that a generic class of tubular nanowires, to be defined in more detail below, are ideal systems for both realizing and observing negative thermopower. 
Semiconductor nanowires are versatile systems with complex phenomenology
attractive for nanoelectronics.  In particular
the thermoelectric current increases due to the lateral confinement
compared to the values in the bulk material \cite{Hicks93}. 
At the same time the thermal conductivity can be strongly suppressed in nanowires
with a diameter below the phonon mean free path \cite{Boukai08,Zhou11}.
These effects together lead to an increased thermoelectric conversion
efficiency in the quasi-one-dimensional geometry.
 
In the tubular nanowires we are interested in, the conduction takes
place only in a narrow shell at the surface, and not through the bulk.
This is realized both in so-called core/shell nanowires (CSNs) 
and topological insulator nanowires (TINs).  
In CSNs this is a consequence of the
structure, the wires being radial heterojunctions of two different
materials, a core and a shell.  When the shell is a conductor and the
core is an insulator, because of the narrow diameter and thickness,
typically 50-100 nm and 5-10 nm, respectively, quantum interference
effects are present, which have been observed as Aharonov-Bohm
magnetoconductance oscillations in longitudinal \cite{Gul14}
and transversal \cite{Heedt16} magnetic fields, and explained with
ballistic transport calculations \cite{Rosdahl14,Rosdahl15,Manolescu16}.  A
tubular conductor can also be achieved with nanowires based on a single
material, but with surface states radially bound due to the pinning of
the Fermi energy \cite{Heedt16}.  In the case of TINs,
the bulk material is an insulator, but with a topologically nontrivial
band structure, that requires a robust metallic state at the surface
\cite{Hasan:2011hs,Bardarson:2013cn}.  Magnetoresistance oscillations,
both in longitudinal and transversal fields, were recently reported 
for TINs made of BiTeSe
~\cite{Bassler15,Peng:2009jm,Xiu:2011hq,Dufouleur:2013bg,Cho:2015gk,Jauregui:2016cx,Dufouleur17}.

In this paper, then, we consider electrons constrained to move on
a cylindrical surface, in the presence of a uniform magnetic field
transversal to the axis of the cylinder, and a longitudinal temperature
bias.  We demonstrate that in these systems a sign reversal of the
thermoelectric current is obtained when varying the magnetic field or the
temperature bias.  Contrary to the cases of molecules and quantum dots,
where the sign change of the current is a result of resonant states, in
these tubular nanowires the effect is a consequence of a nonmonotonic
energy dispersion of electrons vs. momentum.  We further show that the sign reversal
survives in the presence of a moderate concentration of impurities.

The predicted sign reversal of the thermoelectric current
should be detectable in the above mentioned realizations of tubular
conductors, but the magnitude of the anomalous current will
depend on the specific system parameters. 
Considering tubular nanowires of 30 nm radius, infinite length, and
magnetic fields of 2-4 T, we estimate the magnitude of the anomalous
(negative) thermoelectric current as tens of nA.
Thermoelectric transport in CSNs has been already studied in a couple
of experimental papers.  One recent work was the characterization of
GaAs/InAs nanowires by thermovoltage measurements in those situations when
electrical conductance does not provide information \cite{Gluschke15}.
Another study demonstrated enhanced thermoelectric properties in Bi/Te
CSNs via strain effects \cite{Kim2017520}. 

\begin{figure} 
\begin{center}
\vspace{-4mm}
\includegraphics[width=0.40\textwidth]{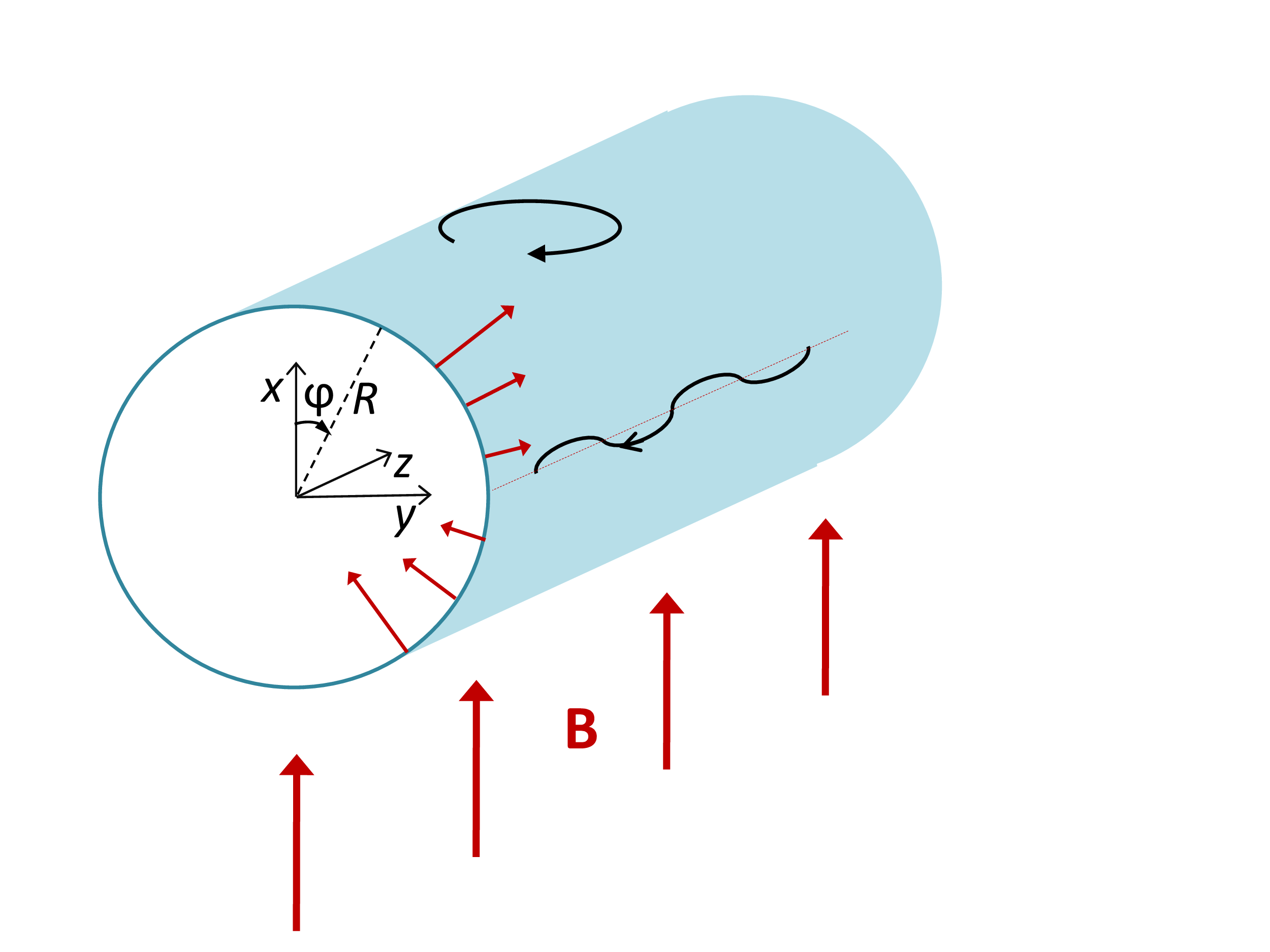} 
\end{center}
\vspace{-6mm}
\caption{A hollow cylindrical nanowire (light blue) in a uniform transverse
magnetic field (thick red arrows).  On the top and bottom regions of the cylinder,
electrons perform closed cyclotronic loops, whereas on the lateral sides longitudinal 
snaking orbits are formed along the lines where the radial
projection of the magnetic field (thin red arrows) is zero.
}
\label{fig:cylinder}
\end{figure}

Electrons constrained to a cylindrical surface, in the presence of a
uniform magnetic field transversal to the axis of the cylinder, have
two types of states: $i$) cyclotron orbits at the top and bottom of the
cylinder, in the direction of the field, where the radial component of
the field is nearly constant, and $ii$) snaking trajectories along the
lateral lines where the radial component vanishes and flips orientation,
such that the Lorentz force always bends the electron trajectory towards
the line \cite{Tserkovnyak06,Ferrari09,Manolescu13,Chang16}; an illustration is
provided in Fig.~\ref{fig:cylinder}.  Such snaking states were studied
earlier in the 90's in a planar electron gas in a perpendicular magnetic
field with alternating sign \cite{Muller92,Ibrahim95,Zwerschke99} and
found responsible for strong positive magnetoresistance in the presence
of ferromagnetic microstrips \cite{Ye95,Manolescu97}.  
For our above-mentioned tubular nanowire the snaking states become ground 
states at nonzero wave vector, imposing a nonmonotonic energy 
dispersion.

%

To focus, we concentrate our detailed discussion on the case of 
CSNs; later we will demonstrate that the effects we 
find are universal and qualitatively the same 
results are obtained for TINs. 
We choose the coordinate system such that magnetic field is along
the $x$ axis, ${\bf B}=(B,0,0)$, the vector potential being ${\bf
A}=(0,0,By)=(0,0,BR\sin\varphi)$.  In this case the 
Hamiltonian can be written as
\begin{equation}
 H=\frac{-\hbar^2}{2m_{\mathrm{eff}}} \!
\left[\frac{\partial^2}{R^2\partial^2 \varphi}+
(\partial_z+\frac{ieBR}{\hbar}\sin \varphi)^2\right] \!
-\frac{g_{\mathrm{eff}}\mu_{\mathrm{B}} }{2} B \sigma .
\label{eq:Hamiltonian}
\end{equation}
In this example we consider material parameters for GaAs, i.e., effective mass
$m_\mathrm{eff}=0.066$ and $g$-factor $g_\mathrm{eff}=-0.44$,
$\mu_{\mathrm{B}}$being the Bohr magneton and $\sigma=\pm 1$ the spin
label.  For $B=0$ the angular part of the Hamiltonian has eigenfunctions
$e^{i \varphi n}/\sqrt{2\pi}$, $n \in \mathbb{Z}$, and the single electron
energies are ordinary parabolas vs. the wave vector $k$ which is defined by the
longitudinal wave functions $e^{ikz}$.  These eigenfunctions define a
basis set, $|nk\sigma\rangle$, which we use for $B\neq 0$ to diagonalize
numerically (\ref{eq:Hamiltonian}). The convergence is reached with
$|n|\leq 50$.

\begin{figure} 
\begin{center}

\vspace{-12mm}
\includegraphics[angle=-90,width=0.60\textwidth]{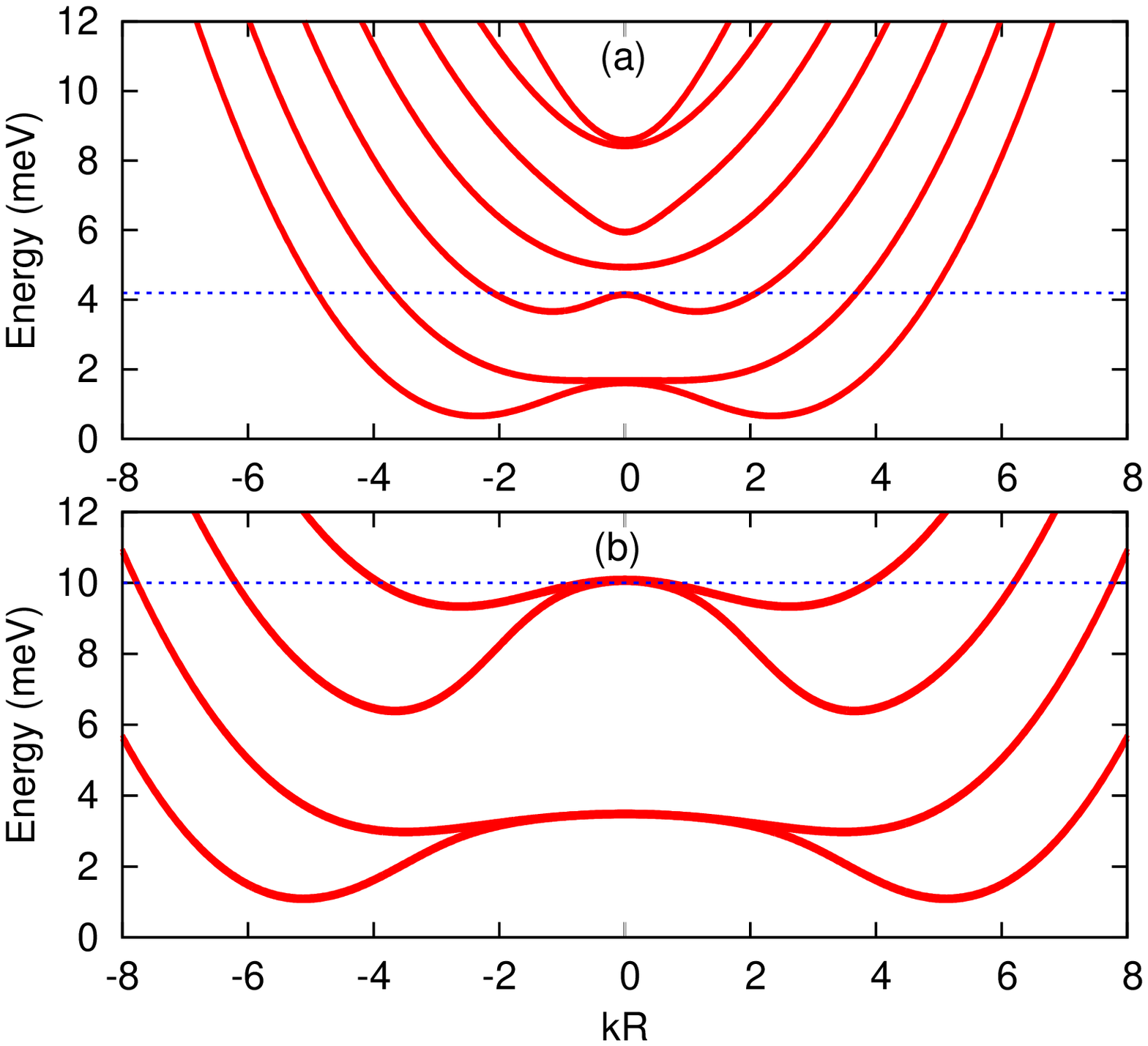}

\end{center}

\vspace{-8mm}
\caption{Energy spectra for a cylinder of infinite length and radius
$R=30$ nm
in a transversal magnetic field $B=2$\,T (a) and $B=4$\,T (b).
The horizontal dotted lines indicate the chemical potential $\mu=4.2$\,meV 
and $\mu=10$\,meV, respectively.
}
\label{fig:energy}
%
%
\begin{center}

\vspace{-5mm}
\includegraphics[angle=-90,width=0.40\textwidth]{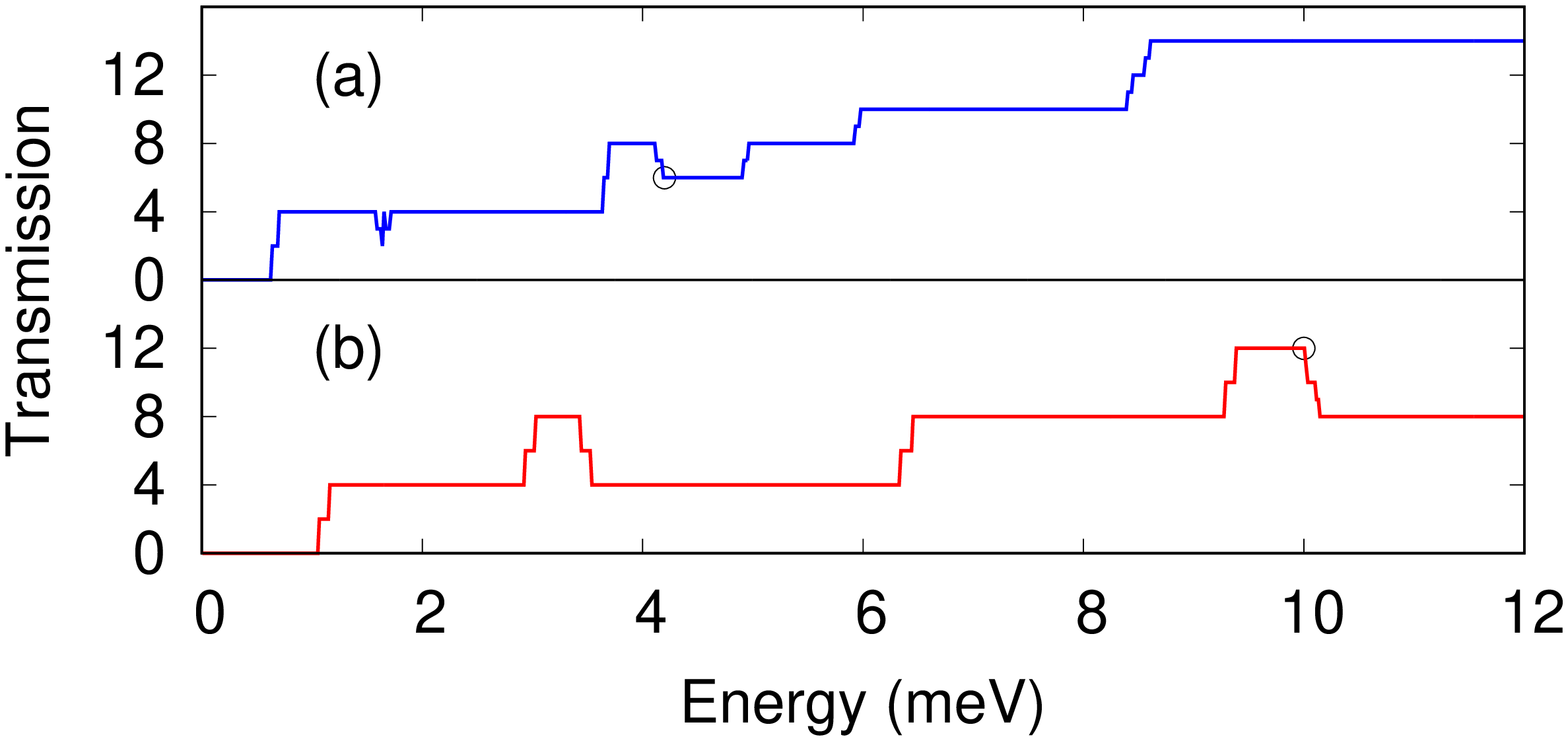}
\end{center}

\vspace{-5mm}
\caption{Transmission function $T(E)$ for $B=2$\,T (a) and $B=4$\,T (b).
The results are derived with the energy spectra shown in Fig. \ref{fig:energy}, 
the circles indicating the location of the chemical potentials.  
The nonmonotonic behavior of the transmission function is clearly seen. 
}
\label{fig:transmission_clean}
\end{figure}

The energy spectra for magnetic fields $B=2.0$\,T and $B=4.0$\,T are
shown in Fig.~\ref{fig:energy}.  
Since the energy of the cyclotron states increases with $B$, at sufficiently 
strong fields the low energy bands have a nonmonotonic
dispersion, with one maximum around $k=0$ and {\it two} lateral symmetric
minima.  The central maximum corresponds to cyclotron orbits (precursors
of Landau levels), and the lateral minima indicate the onset of 
snaking orbits.
At any energy each dispersion curve yields a number of propagating modes.
The usual situation is with one right moving mode, i.e., with $k>0$, for a
given energy.  But for energies lying {\it between} the central maxima and
lateral minima there are two right movers, and accounting for spin results
in four in total.  Because of the very small $g$-factor the spin splitting
is not visible in the figure.  When the energy slightly increases above
the local maximum, one spin pair of propagation modes is excluded.  Hence,
the transmission, which in this case is simply the number of propagating
modes times $e^2/h$, drops two units.  The behavior of the transmission
function $T(E)$ is seen in Fig.~\ref{fig:transmission_clean}, increasing,
but also {\it decreasing}, in steps as a function of energy, as one would
expect from opening and closing modes, respectively. This behavior  
will lead to the sign reversal of the thermoelectric current.

Such a nonmonotonic behavior of the transmission function is also known
for quantum wires with Rashba spin-orbit coupling in a longitudinal 
magnetic field \cite{Nesteroff04,Serra05}.  However, the energy scales 
related to such nonmonotonic
transmission are very small and can only be observed in high quality
cleaved edge overgrowth samples \cite{quay10:336} at temperatures
$\approx 0.3$\,K.  Very recently a similar effect has been observed
in InAs nanowires with a stronger Rashba coupling \cite{Heedt17}.
In contrast, in the present case without spin-orbit coupling, 
the energy scales are much bigger, $T(E)$ is
not smeared out by temperature, and leads to a sign reversal of
the thermoelectric current.

\begin{figure} 
\begin{center}
\includegraphics[angle=-90,width=0.40\textwidth]{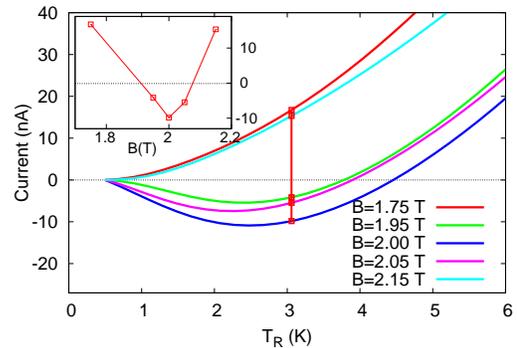}
\end{center}

\vspace{-8mm}
\caption{Thermoelectric current as a function of the temperature of the right 
contact $T_{\rm R}$ for the indicated  magnetic field values and $\mu=4.2$ meV. The inset shows the current as a function of magnetic field for a fixed 
temperature $T_{\rm R}=3$ K marked with a vertical line in the main figure.  
The left contact is kept at constant temperature $T_{\rm L}=0.5$ K. 
 }
\vspace{-5mm}
\label{fig:current_B2}
\end{figure}

The charge current through the nanowire, driven by a temperature
gradient, can be calculated using the Landauer formula
\begin{equation}
I_{\rm c}=\frac{e}{h}\int T(E) \left[f_{\rm R}(E)-f_{\rm L}(E)\right] dE,
\label{eq:Ic}
\end{equation}
where $f_{\rm L/R}(E)$ are the Fermi functions for the left/right reservoir
with chemical potentials $\mu_{\rm L/R}$ and temperatures $T_{\rm L/R}$.
We consider a temperature bias, $T_{\rm R}>T_{\rm L}$, beyond 
linear response, and no potential bias, such that the difference of
the Fermi functions changes sign at $E=\mu_{\rm L}=\mu_{\rm R}$.
Coulomb interactions are neglected, which is a good approximation
for widely open wires.  If the transmission function $T(E)$ increases
with energy over the integration interval the thermoelectric current
is positive, i.e., the electrons flow from the hot contact to the cold one.
This is the normal situation.  An anomalous negative current instead
occurs if the transmission function decreases with energy, as shown in
Fig.~\ref{fig:transmission_clean}.  The energy integral is calculated
numerically using the trapezoidal method.  We keep the left reservoir at
a fixed temperature, $T_{\rm L}=0.5$\,K, i.e., low, but non-zero as in
experimental setups.  By varying the temperature of the right reservoir
we obtain the current as function of $T_{\rm R}$, as shown in Fig.\
\ref{fig:current_B2}, where one can notice that the {\it sign} of the
current may change both with $T_{\rm R}$ or magnetic field.  
%

\begin{figure} 
\begin{center}
\includegraphics[angle=-90,width=0.40\textwidth]{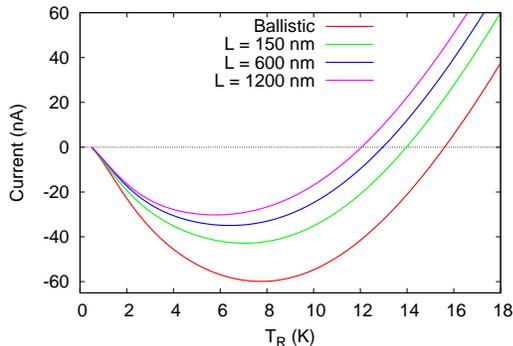}
\end{center}

\vspace{-10mm}
\caption{Thermoelectric current calculated with the energy spectra shown in Fig.\ \ref{fig:energy}
($B=4$ T, $\mu=10$ meV) in the ballistic case and for randomly distributed impurities within 
a scattering region of size $L=150, 600,\ {\rm and}\ 1200$ nm. 
\vspace{-3mm}
}
\label{fig:current_imp}
\end{figure}

The anomalous current can be in the range of tens of nA,
i.e., much larger than for quantum dots.  The largest value shown in
Fig.\ \ref{fig:current_B2} is about -10 nA for $B=2$ T and $T_{\rm R}=2.5$ K.
With a magnetic field of $B=4$ T, yielding the energy spectrum of Fig.\
\ref{fig:energy}(b), we could obtain, in the ballistic case, an 
anomalous current of nearly -60 nA at $T_{\rm R}=8$ K, as 
shown in Fig.\ \ref{fig:current_imp}.

The appearance of the anomalous current relies on nonmonotonic steps
in the transmission function. For clean wires the steps are sharp, but
in the presence of impurities the steps will get rounded.
The transmission function in the case when impurities
are included is obtained using the recursive Green's function method
\footnote{The computational details are described in the Supplemental Material}.
Here we simulate transport in a nanowire 
where the impurities are assumed to be short range,
\begin{eqnarray}
 V_\mathrm{imp}(z,\varphi)=\sum_i W\delta(z-z_i)\delta(\varphi-\varphi_i)\ ,
 \label{eq:Vimp}
\end{eqnarray}
where $W$ is the impurity strength.  
We consider a fixed impurity configuration, i.e., no ensemble average. 
To some extent the results depend on the impurity configuration,
as also seen in experiments. There the conductance can show
complicated, but reproducible behavior for a given nanowire
\footnote{Here reproducible means that the measurement can be repeated
later on the same nanowire and it will give the same conductance
plot.}, whereas the conductance for {\it another} nanowire will yield
conductance whose structure (position of peaks, etc.) will be
different \cite{Wu13:4080}, but reproducible as well.
The average density of impurities is chosen 
$n_\mathrm{imp}=3.0$ nm$^{-1}$
and the disorder strength $W=1.2 \hbar^2/(2m_{\rm{eff}}R^2)$. 
We consider repulsive impurities, $W>0$, since
negative values of $W$ lead to a strong suppression of the conductance when electrons get bound at potential minima.
The key point is that as long as the transmission function still shows 
the nonmonotonic steps the anomalous current is obtained.  
In Fig.~\ref{fig:current_imp} we compare the thermoelectric currents for the same
magnetic field and chemical potential, in the ballistic case
and with a fixed impurity concentration.  
Indeed, the magnitude of the anomalous current is reduced in the
presence of impurities. It further drops for longer wires due to
the increased number of scattering events, but it is still sizeable.
Instead, the magnitude of the normal current increases with the number
of scatterers. This is because the contribution of the transmission bumps
decreases and the transition point $I_{\rm c}=0$ shifts to lower and lower
temperatures.  Of course, if the nanowire is too dirty, such that the
conductance becomes a series of transmission resonances due to quantum
dotlike states \cite{Wu13:4080}, the anomalous current will not be
observable.  However, even in that case the transport calculations
based on elastic scattering reproduced well the thermopower measurements
up to 24 K \cite{Wu13:4080}.  This makes us confident that inelastic
collisions can also be neglected in our temperature range.

\begin{figure} 
\begin{center}

\includegraphics[angle=-90,width=0.40\textwidth]{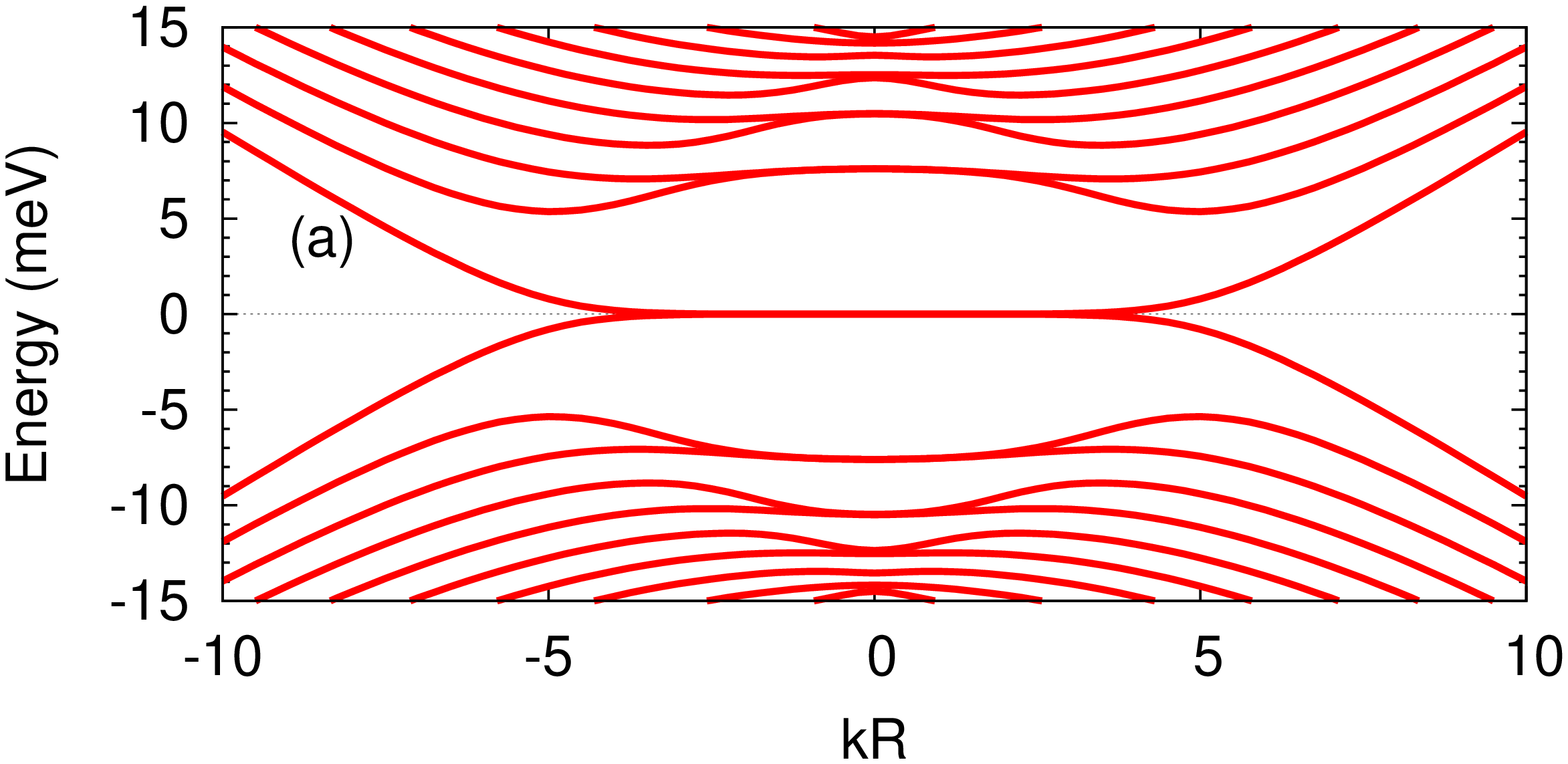}

\vspace{-2mm}
\includegraphics[angle=-90,width=0.40\textwidth]{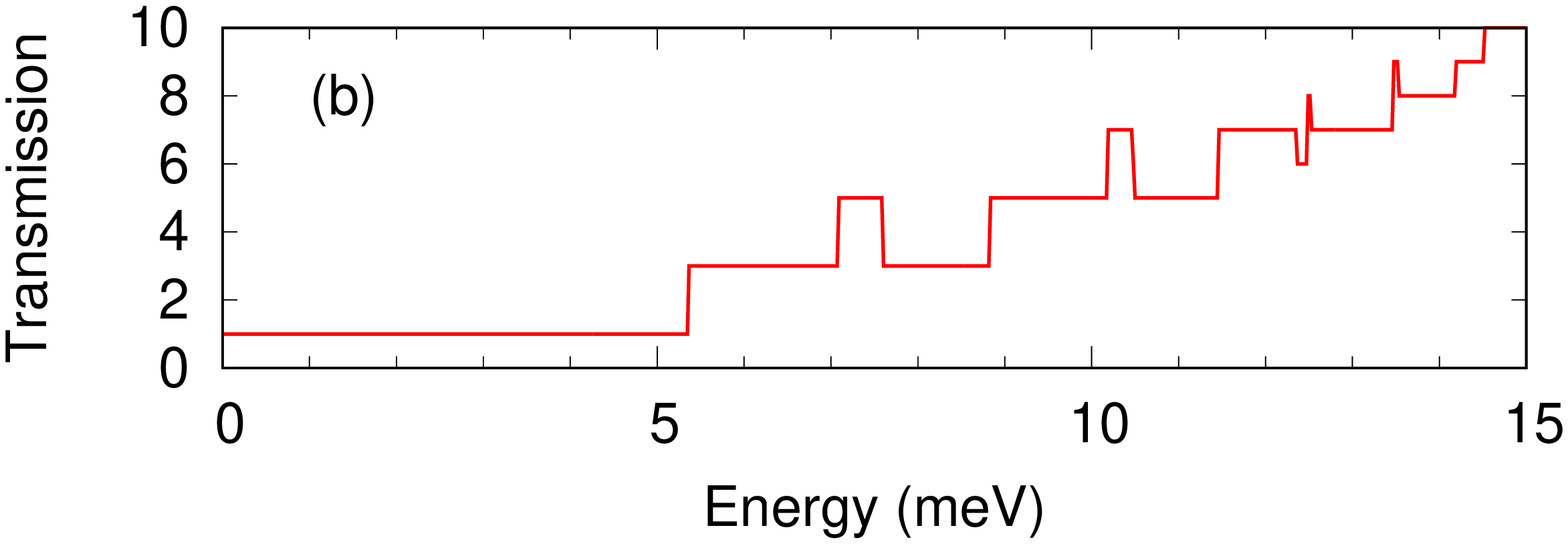}

\vspace{-2mm}
\includegraphics[angle=-90,width=0.40\textwidth]{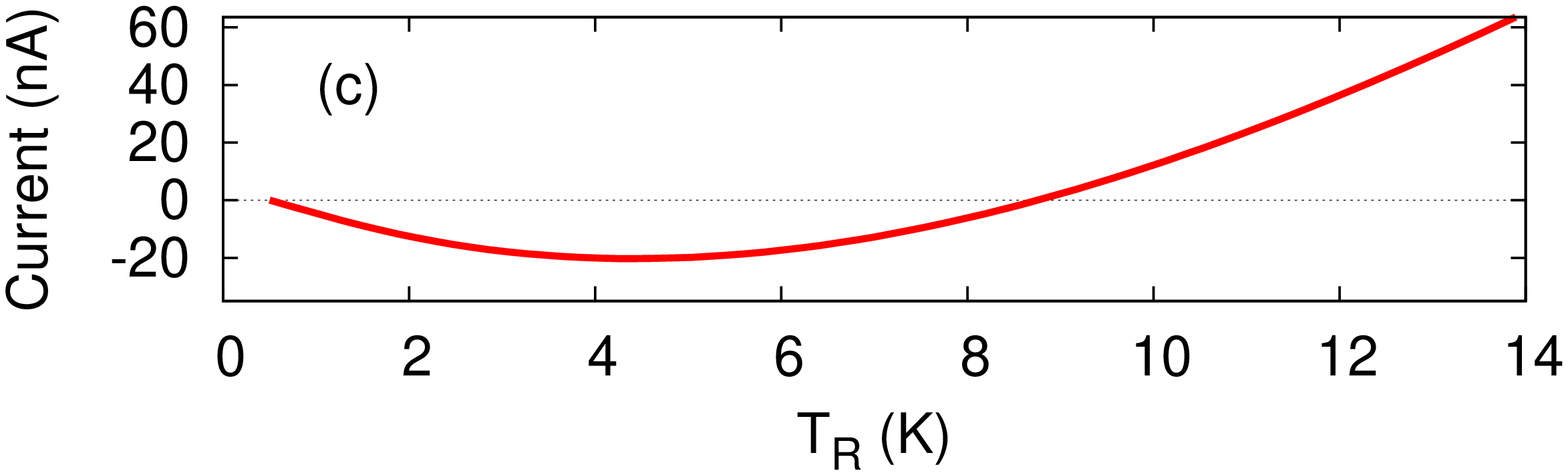}

\end{center}

\vspace{-6mm}
\caption{The energy spectrum (a), transmission function (b), and thermoelectric current (c), 
for a TIN with $v_F=10^5$\,m/s, $R=30$\,nm, and $B=4.3$\,T.
\vspace{-5mm}
}
\label{fig:TInanowire}
\end{figure}

Having considered the CSN case in detail, we now briefly discuss 
the case of TINs.
Such wires in a magnetic field have recently been studied extensively both theoretically~\cite{Bardarson:2010jl,Zhang10:206601,Zhang:2012ci,Ilan:2015ei,Xypakis:2017kw} and experimentally~\cite{Peng:2009jm,Xiu:2011hq,Dufouleur:2013bg,Cho:2015gk,Jauregui:2016cx,Dufouleur17,Arango:2016dd}.
In contrast to the Schr{\"o}dinger fermions of the CSNs, the surface states of the topological insulator are Dirac fermions, described by the Hamiltonian~\cite{Bardarson:2010jl,Zhang10:206601,Bardarson:2013cn}
\begin{equation}
H_\mathrm{TI}=-i\hbar v_F \left[
\sigma_z \left( \partial_z+i\frac{eB}{\hbar}R \sin \varphi \right) +\sigma_y \frac{1}{R}\partial_\varphi
\right ] ,
\label{eq:HamiltonianTI}
\end{equation}
where $v_F$ is the Fermi velocity, and the spinors satisfy antiperiodic boundary conditions $\hat{\psi}(\varphi)=-\hat{\psi}(\varphi+2\pi)$, due to a Berry phase.  
It is convenient to diagonalize (\ref{eq:HamiltonianTI})
using the same angular basis states as before, but because of the boundary condition $n$ now takes half-integer values. 
An example of the energy spectrum is shown in Fig.~\ref{fig:TInanowire}(a) where, as in the CSN case, precursors of Landau levels around $k=0$ are seen, both at negative and positive energy, and snaking states are visible at the edges.
These states give rise to transmission that decreases with energy, as shown in Fig.~\ref{fig:TInanowire}(b), and consequently to an anomalous thermoelectric current, as before, shown now in Fig.~\ref{fig:TInanowire}(c).
The TINs offer some further advances. 
For example, the surface states are robust to disorder, and the negative gradient in the transmission is also obtained at relatively strong disorder strengths.

In conclusion, an unexplored consequence of the coexistence of snaking
and Landau states in tubular nanowires in a transverse magnetic field
is that the transmission function is nonmonotonic with the energy,
which implies that the thermoelectric current can be both positive
and negative.  The normal flow of electrons should be from the hot
to the cold contact.  Instead, in a magnetic field of a few Tesla and
variable temperature of the hot source, here below 10 K, an anomalous
flow occurs, from the cold lead to the hot lead, corresponding to
tens of nA.  This phenomenon can have applications to thermoelectric
devices based on nanowires.  In particular, the detection of the current
reversal can be seen as an indication of the tubular distribution of
the conduction electrons, which is crucial for topological insulator
nanowires.  The presence of snaking states has already been detected
both in CSNs \cite{Heedt16} and in TINs \cite{Bassler15}, and
hence the predicted anomalous current should be within the experimental
reach. Identifying the general relationship between the thermocurrent
and nonmonotonic transmission function can motivate the study of the
anomalous current in other systems.

This work was supported by: RU Fund 815051 TVD, ANCSI Grant PN16420202,
MINECO Grant FIS2014-52564, and ERC Starting Grant 679722. 

\bibliography{thermo.bib}

\onecolumngrid

\vspace{10mm}

\appendix*
\section{Supplemental Material: Computational method for core/shell nanowires with impurities}

\renewcommand{\thefigure}{A\arabic{figure}}
\setcounter{figure}{0}
\setcounter{equation}{0}

\subsection{Green's function discretization}
Our starting point is the partial differential equation for the Green's function of the tubular nanowire
\begin{eqnarray}
\left (
E- \left [
-\frac{\hbar^2}{2mR^2}\partial_\varphi^2+\frac{1}{2m}\left (
\frac{\hbar}{i}\partial_z+eBR\sin\varphi
\right )^2
\right ]
\right )G^r(z,z',\varphi,\varphi';E)=\delta(z-z')\delta(\varphi-\varphi').
\label{eq:GFpde}
\end{eqnarray}
Next, we write $G^r$ as a Fourier series in $(\varphi,\varphi')$ 
\begin{eqnarray}
 G^r(z,z',\varphi,\varphi';E)=\frac{1}{2\pi} \sum_{n,n'\in \mathbb{Z}} e^{-i n\varphi}e^{i n'\varphi'}G^r_{n,n'}(z,z'),
 \label{eq:FourierSeries}
\end{eqnarray}
with Fourier components
\begin{eqnarray}
 G^r_{n,n'}(z,z')=\frac{1}{2\pi} \int d\varphi d\varphi' e^{i n\varphi}e^{i n'\varphi'}G^r(z,z',\varphi,\varphi';E).
\end{eqnarray}
Note that we have dropped the $E$ from $G^r_{n,n'}$ for sake of brevity.  Inserting Eq.\ (\ref{eq:FourierSeries}) into Eq.\ (\ref{eq:GFpde}), multiplying with $e^{-i m\varphi}e^{i m'\varphi'}/2\pi$ and integrating over $(\varphi,\varphi')$ results in
\begin{eqnarray}
\frac{\hbar^2}{2mR^2} \sum_{n\in \mathbb{Z}}\left (
\left (
\mathcal{E}- (n^2-R^2\partial_z^2)
\right )\delta_{m,n}
+2i (R\partial_z)\frac{R^2}{\ell_c^2}[\sin\varphi]_{m,n}
-\frac{R^4}{\ell_c^4}[\sin^2\varphi]_{m,n}
\right )G^r_{n,m'}(z,z')=\delta(z-z')\delta_{m,m'}
\label{eq:GFexpanded}
\end{eqnarray}
Here we introduced $\mathcal{E}$ which measures energy in units of $\frac{\hbar^2}{2mR^2}$ and
\begin{eqnarray}
 [f(\varphi)]_{m,n}=\frac{1}{2\pi} \int d\varphi e^{-i m\varphi}e^{i n\varphi} f(\varphi).
\end{eqnarray}
Equation (\ref{eq:GFexpanded}) can be written more compactly on matrix form
\begin{eqnarray}
\frac{\hbar^2}{2mR^2} \left [
(\mathcal{E}+R^2\partial_z^2)\mathbb{I}- \mathrm{diag}(n^2)
-2i (R\partial_z)\frac{R^2}{\ell_c^2}[\sin\varphi]
-\frac{R^4}{\ell_c^4}[\sin^2\varphi])
\right ]G^r(z,z')=\delta(z-z')\mathbb{I}
\label{eq:GFmatrix}
\end{eqnarray}
Here all matrices are assumed to be of dimension $(2M_\mathrm{max}+1)$, i.e.\ a cut-off has be introduced here such that $n\in [-M_\mathrm{max},M_\mathrm{max}]$.
This is now a equivalent to a one dimensional problem with an internal degree of freedom and one can proceed with standard finite difference implementation where the continuous variable $z$ is discretized.  Introducing a lattice parameter $a$ and the notation $z_{i+1}=z_i+a$, the finite difference version becomes
\begin{eqnarray}
& &\left (E\mathbb{I}-\frac{\hbar^2}{2mR^2}\left [
\mathrm{diag}(n^2)+\frac{R^4}{\ell_c^4}[\sin^2\varphi]+\frac{2R^2}{a^2}\mathbb{I}
\right ] \right ) G^r(z_i,z_j) \nonumber \\
& &
+\frac{\hbar^2}{2mR^2}\left [
-\frac{R^2}{a^2}\mathbb{I}-i \frac{R}{a}\frac{R^2}{\ell_c^2}[\sin\varphi] 
\right ]G^r(z_{i+1},z_j)
+\frac{\hbar^2}{2mR^2}\left [
-\frac{R^2}{a^2}\mathbb{I}+i \frac{R}{a}\frac{R^2}{\ell_c^2}[\sin\varphi] 
\right ]G^r(z_{i-1},z_j)=\delta_{i,j}\mathbb{I},
\label{eq:GF_FDM}
\end{eqnarray}
or in a more compact form
\begin{eqnarray}
\left (E\mathbb{I}-H_\mathrm{sl} \right ) G^r(z_i,z_j)+ V_\mathrm{sl}G^r(z_{i+1},z_j)+ V_\mathrm{sl}^\dagger G^r(z_{i-1},z_j)=\delta_{i,j}\mathbb{I},
\label{eq:GF_FDM_sl}
\end{eqnarray}
where we have defined the slice Hamiltonian
\begin{eqnarray}
H_\mathrm{sl}&=&\frac{\hbar^2}{2mR^2} \left ( \left (\mathrm{diag}(n^2)+ \frac{2R^2}{a^2} \mathbb{I}\right )+\frac{R^4}{\ell_c^4}[\sin^2 ] \right ) ,
\label{eq:Hsl}
\end{eqnarray}
and the slice {\it coupling} Hamiltonian 
\begin{eqnarray}
V_\mathrm{sl}&=&\frac{\hbar^2}{2mR^2} \left ( 
 -\frac{R^2}{a^2} \mathbb{I}+2 i \frac{R^2}{\ell_c^2}\frac{R}{2a}[\sin \varphi ] \right ).
\label{eq:Vsl}
\end{eqnarray}
Note that $V_\mathrm{sl}$ is not hermitian, and does not have to be, since the full Hamiltonian of the entire system contains both $V_\mathrm{sl}$ and $V_\mathrm{sl}^\dagger$.  This is well known in the case of a discretized 2D slab in a non-zero transverse magnetic field
where $V_\mathrm{sl}$ becomes complex via the Peierls substitution \cite{peierls33,ferry97}.

Since Eq.\ (\ref{eq:GF_FDM_sl}) is of the standard hopping or tight-binding type, one can follow standard methods outlined in e.g.\ Ref.\ \onlinecite{ferry97}.  Note that in the absence of magnetic field all the matrices become diagonal and the coupling will simply be determined by the standard tight-binding parameter $t=\frac{\hbar^2}{2ma^2}$.  Physically this means that the eigenmodes $\langle \varphi |n\rangle$ decouple, the different modes are independent and shifted in energy depending on the transverse quantization $\frac{\hbar^2}{2mR^2}n^2$.

\begin{figure} 
\begin{center}
\psfrag{Rth}{ $R \theta$}
\psfrag{zz}{$z$}
\psfrag{SL}{$\Sigma_L$}
\psfrag{SR}{$\Sigma_R$}
\psfrag{H0}{$H_\mathrm{sl}$}
\psfrag{VV}{$V_\mathrm{sl}$}
\psfrag{VD}{$V_\mathrm{sl}^\dagger$}
\psfrag{Hi}{\!\!\!\! $H_i$}
\psfrag{dots}{$\dots$}
\psfrag{ii}{$z_i$}
\psfrag{i1}{\!\!\!$z_1$}
\psfrag{iN}{\!\!\!$z_N$}
\includegraphics[width=0.70\textwidth]{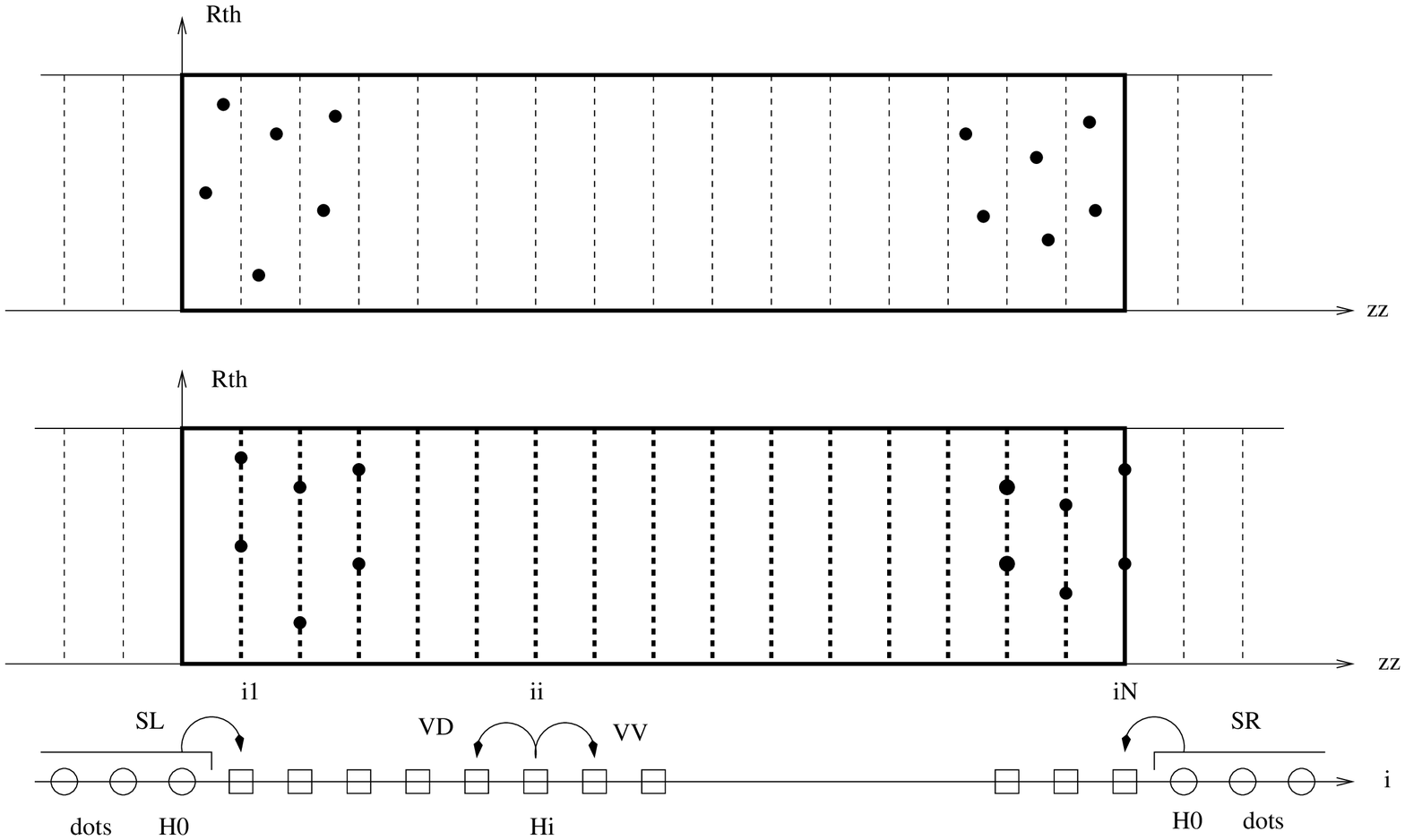} 
\label{fig:sys1}
\end{center}
\caption{The tubular nanowire represented as a rolled out rectangular sheet.  The top figures shows the impurities distributed randomly in the central part.  The middle figure shows the impurities after they are shifted to a discrete value $z_i$, depending on their initial position.  The bottom figures shows a one-dimensional tight-binding model with slice matrices $H_i$ in the central part and coupling matrices $V_\mathrm{sl}$.  The left (right) infinite parts gives rise to self-energies $\Sigma_L$ ($\Sigma_R$) that inter $H_1$ ($H_N$).}
\end{figure}

To describe a wire in the quasi-ballistic regime the discrete values $z_i$ are separated into left, central, and right part, see Fig.\ \ref{fig:sys1}. The impurities only occur in the central part, where the short range impurity Hamiltonian (Eq.\ (3) in the manuscript) is discretized leading to a new Hamiltonian for slice $i$
\begin{eqnarray}
 H_i=H_\mathrm{sl}+V_{\mathrm{imp},i}.
\end{eqnarray}
In the left and right parts of the system all the slice Hamiltonians, and slice couplings, are the same, $H_\mathrm{sl}$ and $V_\mathrm{sl}$ respectively, the self-energies
can be calculated using a very fast, versatile algorithm \cite{LopezSancho85:851}.  Note that the self-energies only enter the slice Hamiltonians for slice $i=1$ ($\Sigma_L$) and $i=N$ ($\Sigma_R$).   Once the self-energies are found, as a function of energy the Green's function in the central region can be found using the recursive Green's
function method \cite{ferry97}. 
From the recursive algorithm one obtains $G^r_{z_1,z_N}(E)$, the part of the Green's function that describes the connection of the first slice (1) and the last slice ($N$), see bottom of Fig.\ \ref{fig:sys1}.  Finally, after the Green's function is obtained
the transmission function can found using the Fisher-Lee relation \cite{Fisher81:6851}.
\begin{eqnarray}
 T(E)=\mathrm{tr}\{\hat{\Gamma}_L(E)G^r(z_1,z_N)(E)\hat{\Gamma}_{R}(E)(G^r(z_1,z_N)(E))^\dagger 
 \},
\end{eqnarray}
where $\hat{\Gamma}_\alpha(E)=-i(\hat{\Sigma}_\alpha(E)-(\hat{\Sigma}_\alpha(E))^\dagger)$ for $\alpha=L,R$.  

\subsection{Choice of lattice parameter $a$}
In zero magnetic field the lattice version of the Green's function, Eq.\ (\ref{eq:GF_FDM}) is diagonal in the angular basis, so the system consists of independent one-dimensional modes.  The dispersion of a nearest neighbor tight binding system with coupling $t=\frac{\hbar^2}{2ma}$ is
\begin{eqnarray}
 \varepsilon(k)&=&2t (1-\cos(ka)).
 \label{eq:cosine}
\end{eqnarray}
To use the discretization as an approximation to the continuum version, the value of $a$ needs to be small enough such that Eq.\ (\ref{eq:cosine}) is close enough to the correct parabolic dispersion up to some relevant maximum value of $k$.  A natural choice for the maximum value is $ k_M=\tfrac{M}{R}$
which corresponds to having parabolic dispersion up to transverse state $M$ with energy 
$\tfrac{\hbar^2}{2mR^2}M^2$.
To quantify the relative deviation of the cosine dispersion from the parabolic one, we use
\begin{eqnarray}
\left . \frac{
\frac{\hbar^2k^2}{2m}-\varepsilon(k)
}
 {
 \frac{\hbar^2k^2}{2m}
 } \right |_{k=\frac{M}{R}} 
 =
1-\frac{1}{M^2}\frac{R^2}{a^2}
2 \left (1-
\cos \left (
M\frac{a}{R} \right ) \right )<\eta,
 \end{eqnarray}
 where $\eta$ is the required relative accuracy.  Now, writing $a=s\tfrac{R}{M}$, where $s$ is dimensionless quantity, we have 
\begin{eqnarray}
1-\frac{2}{s^2}
\left (1-
\cos \left (
s \right ) \right )<\eta.
 \end{eqnarray}
For a given value of $\eta$, e.g.\ $\eta=0.01$, one obtains that $s<0.35$ will give a $1\%$ relative accuracy.

When a non-zero magnetic field is considered, this procedure does not strictly hold.  However, when a confined system with typical size $R$ is subjected to an external magnetic field the relevant length scale  becomes the magnetic length $\ell_c$, for large enough magnetic fields.  To reflect this cross-over of length scales we use
\begin{eqnarray}
 a&=&0.35 \frac{R}{M} \left (1+\frac{R^4}{\ell_c^4} \right )^{-1/4}.
\end{eqnarray}
As a final check we do check that numerical results (energy spectrum and transmission function) does not change significantly for smaller values of $a$.

\end{document}